\documentclass[preprint,onecolumn]{aastex}

\usepackage{bm}
\newcommand{\apx}[1]{^{\mbox{\tiny{(#1)}}}}

\begin{document}

\title{Magnetic Diagnostics of the Solar Chromosphere with the 
\ion{Mg}{2} h--k Lines}

\author{T.\ del Pino Alem\'an,$^a$ R.\ Casini,$^a$ and R.\ Manso Sainz$^b$}

\affil{$^a$High Altitude Observatory, National Center for Atmospheric
Research,\footnote{The National Center for Atmospheric Research is sponsored
by the National Science Foundation.}\break
P.O.~Box 3000, Boulder, CO 80307-3000, U.S.A.\break
$^b$Max-Planck-Institut f\"ur Sonnensystemforschung,\break
Justus-von-Liebig-Weg 3, 37077 G\"ottingen, Germany}

\begin{abstract}
We investigated the formation of the \ion{Mg}{2} h--k doublet in a weakly 
magnetized atmosphere (20--100\,G) using a newly developed numerical code 
for polarized RT in a plane-parallel geometry, 
which implements a recent formulation of partially coherent scattering by
polarized multi-term atoms in arbitrary magnetic field regimes. 
Our results confirm the importance of partial redistribution effects in 
the formation of the \ion{Mg}{2} h and k lines, as pointed out by 
previous work in the non-magnetic case. We show that the 
presence of a magnetic field can produce measurable
modifications of the broadband linear polarization even for relatively small
field strengths (${\sim}10$\,G), while the circular polarization remains
well represented by 
the classical magnetograph formula. Both these results open an important
new window for the 
weak-field diagnostics of the upper solar atmosphere.
\end{abstract}

\maketitle

\section{Introduction}
\label{sec:intro}

One of the big challenges faced these days by the solar physics community 
is to gain a solid understanding of the solar chromosphere, 
and how it magnetically connects to the underlying photosphere and the
corona above.

Within the chromosphere, the structure and dynamics of the magnetized 
plasma undergo dramatic changes. This region spans approximately nine 
pressure scale heights, and the gas temperature goes through a minimum 
of only a few thousand K, before suddenly rising to the million K 
temperatures of the solar corona.  
As the gas density decreases, the intrinsic three-dimensional 
distribution of the solar radiation becomes increasingly important, 
because the excitation of the chromospheric ions becomes more 
strongly correlated with the degree of anisotropy of the radiation. 
At the same time, the reduced role of particle collisions in 
thermalizing the atomic populations allows for subtle quantum 
effects (e.g., atomic polarization, level-crossing coherence, the 
magnetic and electric Hanle effects in the presence of deterministic 
as well as turbulent fields) to become apparent in the spectral and 
polarization signatures of the solar chromosphere \citep{TB01,LL04,CL08}. 

\begin{figure*}[!t]
\centering
\includegraphics[width=.495\hsize]{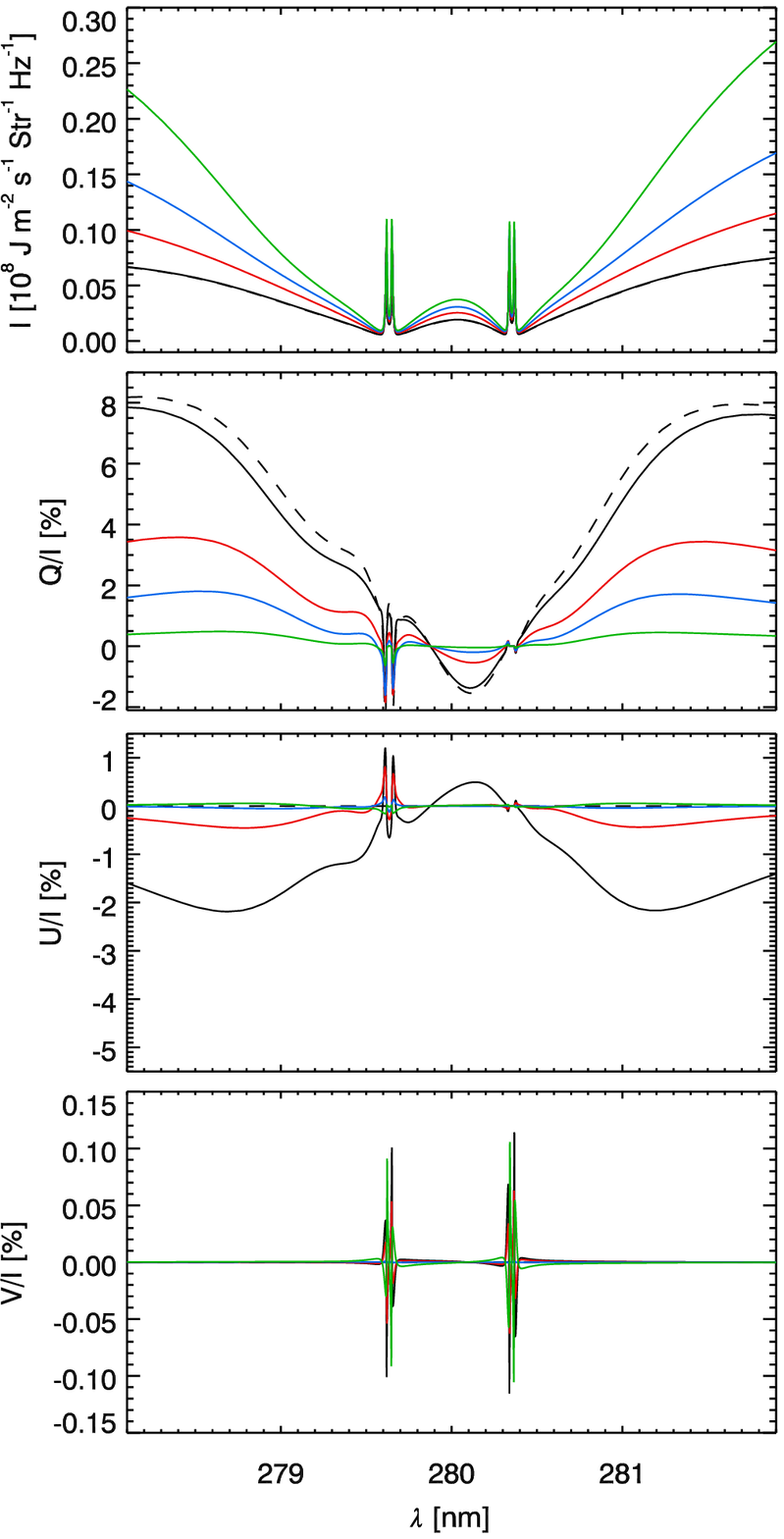}
\includegraphics[width=.495\hsize]{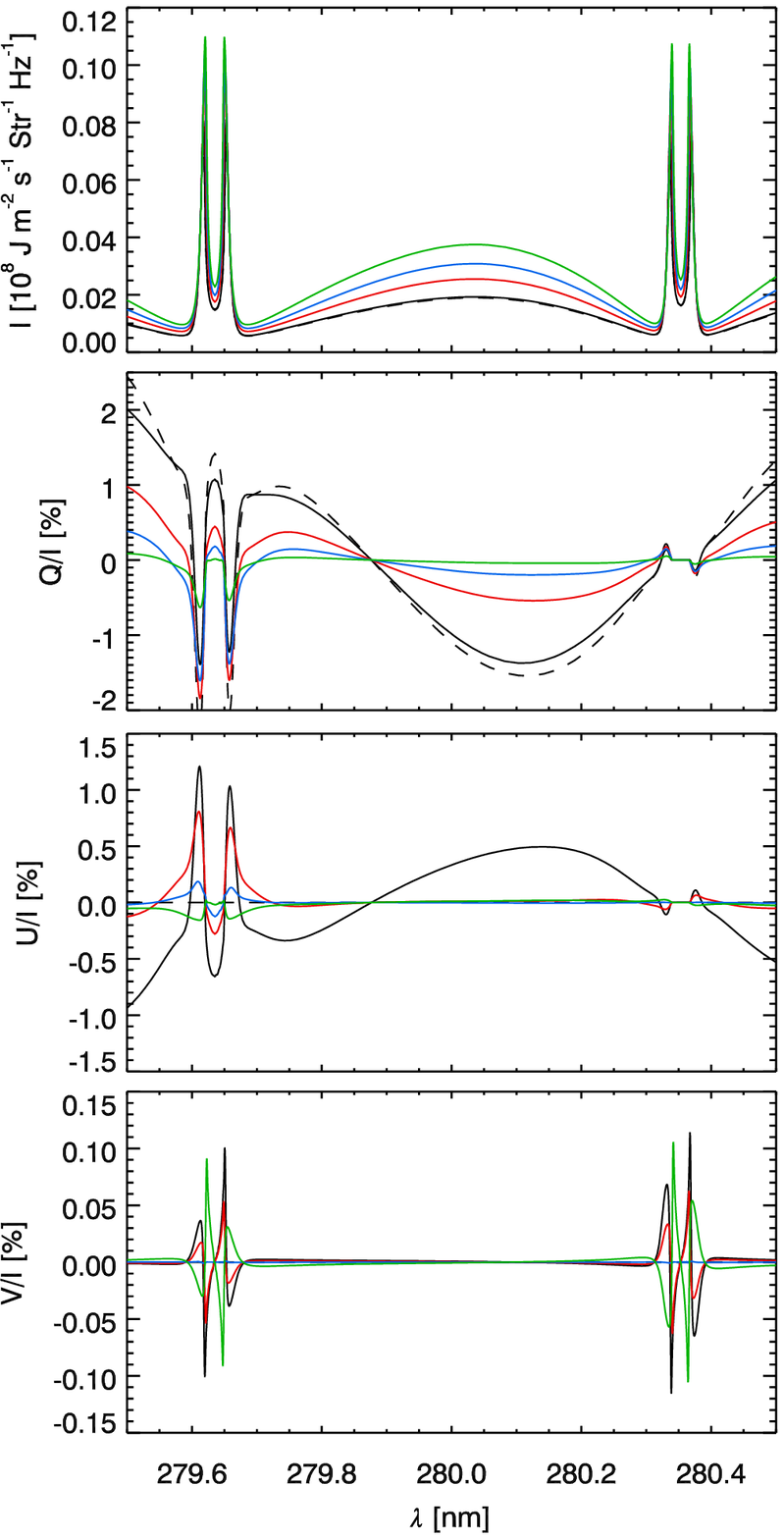}
\caption{\label{fig:both}
Stokes profiles of the \ion{Mg}{2} h--k doublet modeled in a weakly
magnetized FAL-C atmosphere ($B=20$\,G, $\vartheta_B=30^\circ$, 
$\varphi_B=180^\circ$) and for various directions of the LOS 
(corresponding to $\mu=0.1,0.3,0.5,0.8$, respectively, for the 
black, red, blue, and green curves). The dashed curves correspond
to the Stokes profiles for the non-magnetic case and $\mu=0.1$.
\emph{Left:} note the remarkable presence of broadband Stokes-$U$ 
polarization due to the combination of upper-term quantum interferences 
and magneto-optical effects. \emph{Right:} finer details of the polarization 
of the h and k line cores and of the quantum interference pattern between 
them. We note in particular the reversal of the sign of Stokes $V$ (and 
more subtly, of Stokes $U$) for $\mu=0.8$, in accordance with the sign of 
the LOS projection of the magnetic field vector. We also note the complete
absence of a magnetic signature in the core of the h-line at 280.35\,nm,
as expected for an intrinsically non-polarizable transition in the
weak-field limit.
}
\end{figure*}

\begin{figure*}[!t]
\centering
\includegraphics[width=.495\hsize]{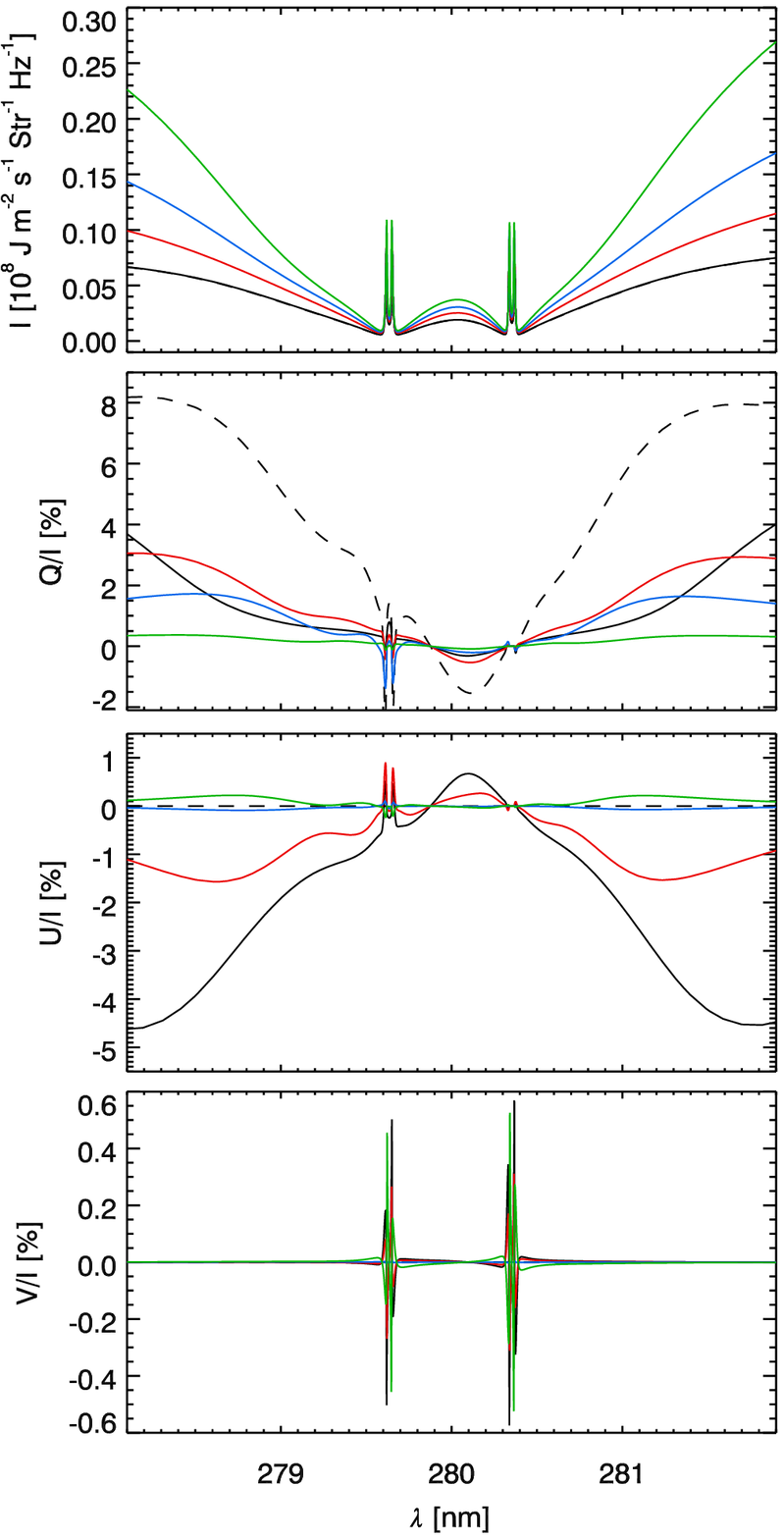}
\includegraphics[width=.495\hsize]{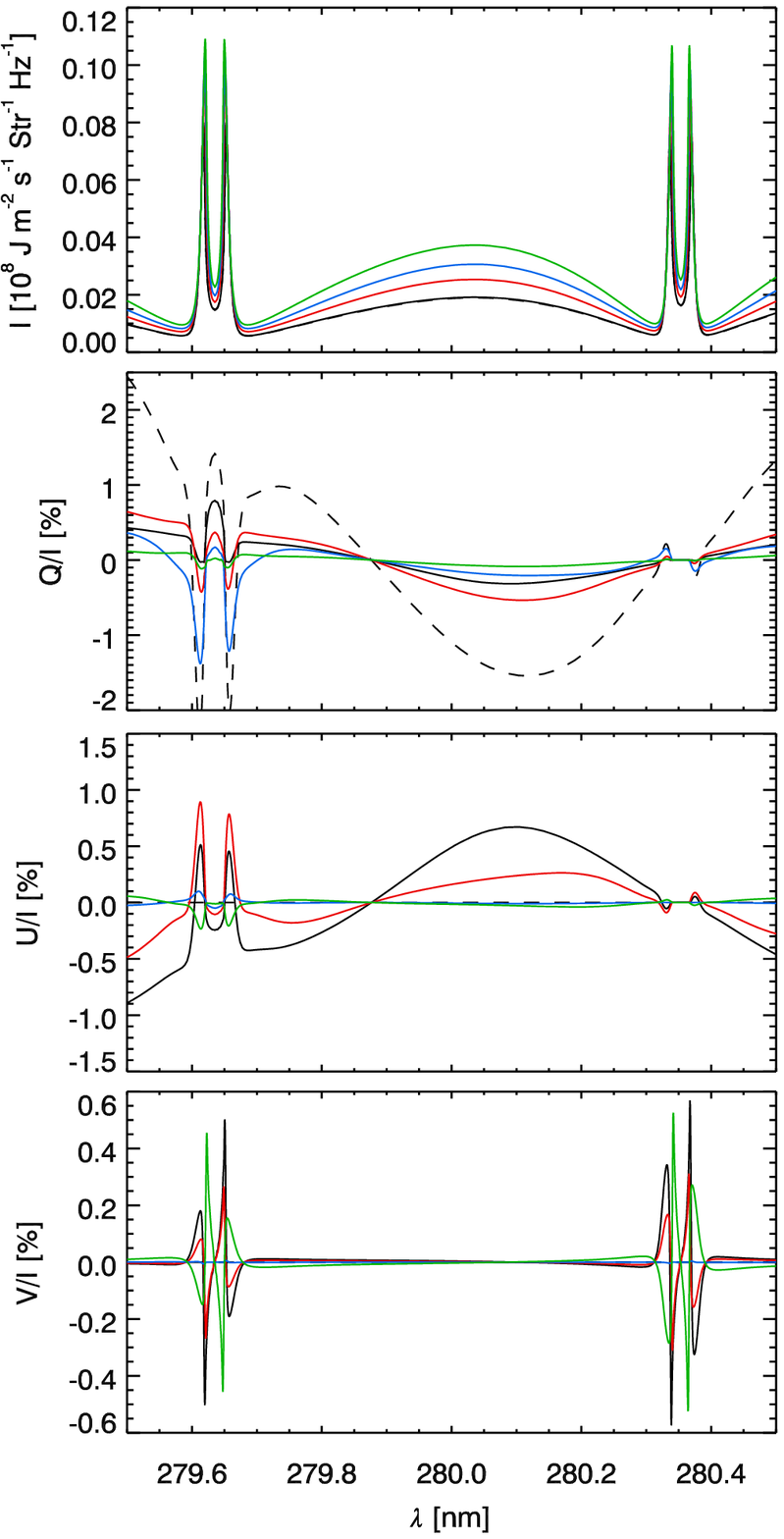}
\caption{\label{fig:both.100}
Same as Figure~\ref{fig:both}, but for a magnetic field strength
$B=100$\,G. We note how the separation of the lobes in the broadband 
polarization structure of Stokes $Q$ and $U$ increases with the magnetic
strength because of the M-O effects. The changes in the line cores are
instead dominated by the depolarization associated with the larger field
strength.
}
\end{figure*}

\begin{figure*}[!t]
\centering
\includegraphics[width=.495\hsize]{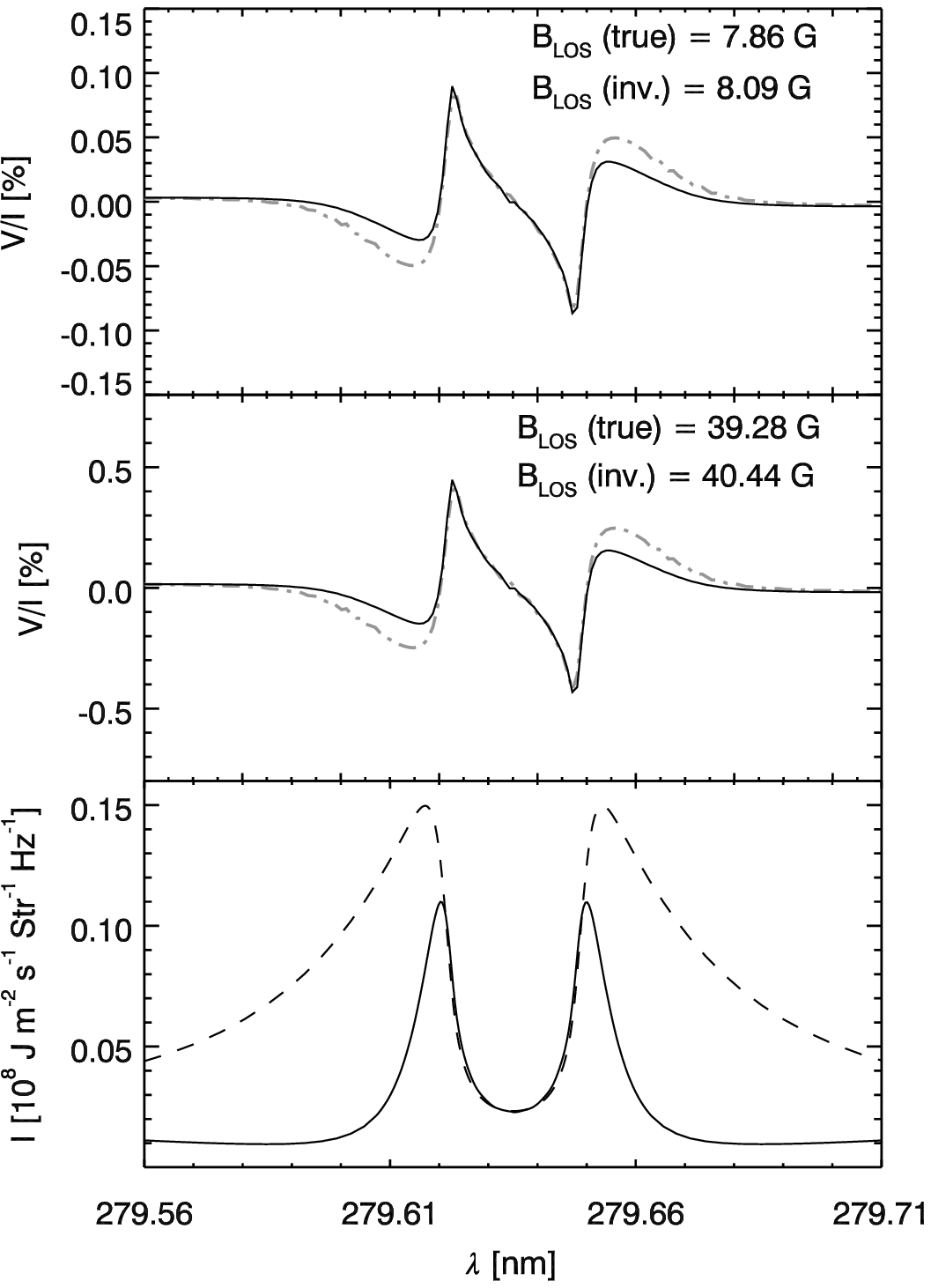}
\caption{\label{fig:only_k}
Fractional circular-polarization profiles $v = V/I$ of the \ion{Mg}{2} 
k-line, for the magnetic models of Figure~\ref{fig:both} (top) and 
Figure~\ref{fig:both.100} (center), and corresponding intensity profiles 
(bottom) for the CRD (dashed) and PRD (solid) regimes. Only the case of 
$\mu=0.8$ is shown here (cf.\ green curves of Figures~\ref{fig:both} 
and \ref{fig:both.100}), showing the fully resolved line core and the
near wings of the line. 
The dashed-dotted curves show the weak-field approximation to 
Stokes $V$ (magnetograph formula). We note the extremely good fit of
this approximation in the spectral region where the line formation
is dominated by the CRD regime. The reported values of the inverted LOS 
field component were estimated by restricting the use of the magnetograph
formula to the inner lobes of the k-line, where the $v$ fractional 
polarization is larger than 0.04\% and 0.2\%, respectively,
for the 20\,G and 100\,G field strengths.
}
\end{figure*}

Another increasingly important aspect of the modeling of 
chromospheric spectral lines in realistic solar scenarios is the ability 
to account for the higher temporal coherence between the processes of 
absorption and re-emission of the solar radiation, which is 
fostered by the particular physical state of the tenuous chromospheric 
plasma. This condition of \emph{partially coherent scattering}
gives rise to a plethora of phenomena (commonly dubbed \emph{partial 
frequency redistribution} or PRD), 
which must be taken into account 
for a proper diagnosis of the plasma and magnetic properties of
the chromosphere.
In particular, PRD effects are fundamental for the interpretation 
of many quantum interference patterns that are observed in the
solar spectrum even between widely separated multiplet lines. This 
was originally demonstrated by \cite{St80} in the case of the 
\ion{Ca}{2} H--K doublet around 395\,nm. Such effects are thus 
essential also for the modeling of the linear polarization of the 
\ion{Mg}{2} h--k doublet around 280\,nm \citep{Au80,HS87,BT12}, 
which has a quantum structure similar to that of the
\ion{Ca}{2} H and K lines.
Recently, polarized radiative transfer (RT) with PRD in the non-magnetic 
case has been applied to the modeling of a variety of chromospheric 
line multiplets showing quantum interferences \citep{Sm12,BT14},
as well as in the case of a uniformly magnetized slab \citep{Sm13}.

Independently, \citeauthor{Ca14}\ (\citeyear{Ca14}; see also 
\citeauthor{CM16} \citeyear{CM16}) have attacked the problem 
of the formation of spectral line polarization by partially coherent 
scattering in a magnetized medium, with sufficient generality to enable 
the modeling of many resonance lines of the solar spectrum that show 
complex linear polarization patterns 
\citep{Wi75,SK96,SK97,St00,Ga00}. In particular, that formalism allows 
to fully take into account the role of atomic polarization in the lower 
state of an atomic transition, a feature that has been neglected by 
previous works in PRD modeling. On the other hand, lower-level 
polarization is important for the interpretation of many 
chromospheric diagnostics, as demonstrated by \cite{MT03} in the 
case of the \ion{Ca}{2} IR triplet.

In order to apply the formalism of \cite{Ca14} under realistic 
chromospheric conditions, we have developed a 1-D RT code for the polarized multi-term atom 
in an arbitrary magnetic field,
which takes into account the effects of PRD, 
as well as the contribution of (isotropic) inelastic and elastic collisions. 
The code is based on a straightforward $\Lambda$-iteration scheme
\citep[e.g.,][]{Mi78}, and it arrives at the solution for the polarized 
PRD transfer problem in a magnetized atmosphere in two steps. 
In the first stage we assume \emph{complete frequency redistribution} (CRD), 
and solve the non-LTE problem of the second kind \citep{LL04} for zero
magnetic field and including only inelastic collisions.
In order to facilitate the convergence of the CRD problem, we 
initialize the level populations with the non-LTE solution from the 
RH code \citep{Ui01}.
In the second stage, this converged CRD solution is
used to initialize the iteration for the magnetized PRD problem, with the 
further addition of elastic collisions. The effects of collisions are 
taken into account in the RT equation by implementing 
physically consistent branching ratios between the first-order (CRD) 
and second-order (PRD) emissivity terms of the theory \citep[cf.][]{Ca14}.
As a result, the emissivity term in the RT equation takes the form
\begin{equation} \label{eq:emissivity}
\varepsilon_i(\omega_k,\hat{\bm{k}})=
\left[\varepsilon_i\apx{1}(\omega_k,\hat{\bm{k}})
-\varepsilon_i\apx{2}(\omega_k,\hat{\bm{k}})_{\rm f.s.}\right]
+\varepsilon_i\apx{2}(\omega_k,\hat{\bm{k}})
\end{equation}
where $\varepsilon_i\apx{2}(\omega_k,\hat{\bm{k}})_{\rm f.s.}$ corresponds
to the expression of $\varepsilon_i\apx{2}(\omega_k,\hat{\bm{k}})$ in the
limit of flat-spectrum (f.s.) illumination. Using equation~(15) of
\cite{Ca14}, it is straightforward to verify that the emissivity
(\ref{eq:emissivity}) converges to the one of \cite{Ui01}, in
the case of the unpolarized multi-level atom.

\section{Model and Results}
\label{sec:description}

We used our new code to study the formation of the Stokes
profiles of the
\ion{Mg}{2} h and k lines (respectively at 280.4\,nm and 279.6\,nm) 
in a homogeneously magnetized atmosphere. For zero
magnetic field, we verified that the results of our code for the
linear polarization of these lines agree with those
presented by \cite{BT12,BT14}.

Figure~\ref{fig:both} shows a synthesis of the four Stokes profiles 
(intensity $I$, fractional linear polarization $q = Q/I$ and $u = U/I$, 
fractional circular
polarization $v = V/I$) of the \ion{Mg}{2} h--k doublet formed in a 
\emph{uniformly magnetized} chromosphere, observed 
on-disk for values of $\mu=0.1,0.3,0.5,0.8$ (respectivily, black, red, blue, and green curves). 
We used the FAL-C atmospheric model \citep{FAL93}, according to 
which these lines form 
in a layer spanning approximately between 200\,km (far wings) to 
2200\,km (line core) above the photosphere. For the magnetic 
modeling, we assumed a field with strength $B=20$\,G, 
inclined $30^\circ$ from the local vertical, and with its
projection on the plane of the sky 
pointing radially toward the solar limb (i.e., $\varphi_B=180^\circ$).
We note that the field strength of 20\,G corresponds
approximately to the critical Hanle field for the \ion{Mg}{2}
k line \citep{BT11}.
We also note that for $\mu=0.5$ (blue curve) the field is perfectly 
transversal to the line-of-sight (LOS), a condition that causes 
Stokes $V$ to vanish.

The left panels of Figure~\ref{fig:both} show 
the line cores and the region between the two lines with increased
spectral details.
We see how the core of the k-line Stokes $Q$ and $U$ in the magnetic case
is affected by the presence of the weak magnetic field through the Hanle
effect, showing a depolarization of the Stokes $Q$ signal (with respect
to the case of zero magnetic field, shown here only for the case
of $\mu=0.1$; see black dashed curves) and the
corresponding appearance of a signal in the core of Stokes $U$. 
\emph{In an optically thin plasma, and in the CRD limit, these
would be the only observable effects in the linear polarization profiles of
this line.} Because the h-line is intrinsically unpolarizable by
radiation anisotropy \cite[being a 1/2--1/2 transition;][]{Ca02}, no 
linear polarization is detected in its core, since the ambient 
magnetic field is too weak to produce any significant Zeeman-effect 
signal. For the same reason, being the Larmor frequency much 
smaller than the fine-structure separation of the upper term, there are 
also no detectable effects of \emph{atomic orientation} (i.e., the 
magnetically induced population imbalance between atomic states of 
the form $(J,\pm M)$; \citealt{Ke84}) in the circular polarization of 
the two lines.

The inclusion of PRD effects in the modeling is essential in order to
produce the broadband linear polarization structure of the 
\ion{Mg}{2} h--k doublet, as already demonstrated by \cite{BT12}
for Stokes $Q$ in the non-magnetic case. In particular, this 
polarization structure reveals the presence of quantum interferences 
between the $J=1/2$ and $J=3/2$ states of the upper term of the doublet. 
A striking result of our modeling is the fact that a corresponding 
broadband linear polarization manifests also in Stokes $U$ (see, e.g., 
Figure~\ref{fig:both}), in the presence of a magnetic field, rather than 
only in the core as one would have expected based on the Hanle-effect mechanism.

An in-depth analysis of the line formation problem demonstrates
that the appearance of this polarization in the broad wings of Stokes-$U$
is mainly caused by the manifestation of
magneto-optical (M-O) effects in the optically thick
chromosphere. These effects are able to transform the 
broadband polarization that is observed in Stokes $Q$ (even in the 
absence of a magnetic field) into detectable levels of Stokes-$U$ 
polarization, already for magnetic fields of only a few gauss.
Our analysis has also showed that a smaller broadband modulation of the 
Stokes-$U$ polarization signal is additionally produced by the breaking of 
the cylindrical symmetry of the scattered radiation, as this is transported 
across the optically thick and magnetized plasma, despite the
restriction of the model to a plane-parallel atmosphere, and even 
when M-O effects are neglected. This is demonstrated by the
case of $\mu=0.5$ (blue curve), where the magnetic field becomes 
exactly perpendicular to the LOS, thus making the M-O element 
responsible for the $Q\,{\to}\,U$ transfer identically zero.

Figure~\ref{fig:both.100} reproduces the
previous results in the case of a magnetic field with $B=100$\,G and the
same geometry as before. We must note that, for this field strength, 
the Hanle effect of the k-line is practically saturated. The figure
shows how the larger magnetic
strength affects the amplitude and separation of the polarization 
lobes observed in the broadband structure of Stokes $Q$ and $U$
outside of the h--k spectral
range. This is clearly a M-O effect, since the associated dispersion
profile is sensitive to the Zeeman splitting of the atomic levels
induced by the presence of the ambient magnetic field. In the
core of Stokes $Q$ and $U$, instead, the changes are dominated by the
depolarization of the signal in the saturation regime of the Hanle effect.
Stokes $V$ is evidently dominated by the longitudinal Zeeman effect, and
so its amplitude scales linearly with the magnetic strength.

An analysis of the Stokes-$U$ contribution to the source term
in the RT equation as a function of height reveals that 
these M-O effects in the very far wings of the h--k doublet occur in 
the upper photosphere, at a height of ${\sim}200-500$\,km in the FAL-C model 
atmosphere adopted for these simulations. This is true also for the 
spectral region between the two lines. The line core polarization is
instead produced in the low transition region (above ${\sim}2000$\,km).

The top and center panels of Figure~\ref{fig:only_k} show the details 
of Stokes $V/I$ in the spectral range around the \ion{Mg}{2} k-line,
for the models of Figures~\ref{fig:both} and \ref{fig:both.100}, 
respectively, and in the case of $\mu=0.8$ (black solid curve).
Overplotted with the gray dashed-dotted curve is the \emph{weak-field 
approximation} of the circular-polarization profile. 
For our model atmosphere with a uniform magnetic field, this
turns out to be proportional to the first derivative of Stokes $I$ 
with respect to wavelength, the so-called \emph{magnetograph 
formula} \cite[e.g.,][]{LL04}.
For comparison, the bottom panel shows the intensity profile of the 
k-line (black solid curve). Overplotted with the gray curve is the 
same profile in the CRD limit.

We note how the weak-field approximation fails to 
quantitatively reproduce the secondary lobes of the Stokes-$V$ profile. 
These are formed in the outer wings of the characteristic peaks of the 
\ion{Mg}{2} k intensity profile, and are therefore sensible to the 
presence of PRD effects in the line forming region (see bottom panel). 
Conversely, the magnetograph formula appears to work extremely well over 
a spectral region that extends from the line center out to peaks of the 
intensity profile, where the CRD approximation to the line formation is
applicable. With this restriction, our results indicate that the 
magnetograph formula retains its diagnostic value, even in complex 
atmospheric scenarios as the one described by our model.

An analysis of the contribution function for Stokes $V$ shows
that the secondary lobes are formed in the chromosphere proper 
(between approximately 1000 and 2000\,km). 
Hence, \emph{the magnetograph formula applied to the \ion{Mg}{2} h--k 
doublet provides a valuable quick-look magnetic diagnostics of the low 
transition region, whereas it may significantly underestimate the 
strength of chromospheric fields.}

In particular, the inverted
values for the LOS strength reported in Figure~\ref{fig:only_k} were 
estimated by applying the magnetograph formula in a spectral range 
where the fractional circular polarization is 
larger than 0.04\% (0.2\%) for the case of $B=20$\,G (100\,G). In 
that restricted region, the relative error on the inverted field 
strength caused by the weak-field approximation is only a few percent.

We must also point out that, for
a given optical depth, observations closer to disk center (i.e., for
larger values of $\mu$) probe larger geometric depths in the
atmosphere, and so we can expect that the error in the application of
the magnetograph formula would increase in the presence of
magnetic-field gradients along the atmospheric height. On the
other hand, the validity of the weak-field approximation in the CRD
spectral region of these lines must remain true \emph{at each height} 
of the atmospheric model, potentially offering a computationally
economic tool for the forward modeling of Stokes $V$ in the presence of
atmospheric gradients. This conclusion is supported by previous 
findings about the applicability of bisector methods to infer the height 
dependence of the magnetic field in chromospheric lines 
\cite[e.g.,][]{Ui03,Wo10}.

\section{Conclusions}
\label{sec:concl}

The following conclusions can be drawn from the above results.
First of all, the importance of PRD effects for the formation of the 
linear polarization profiles of the \ion{Mg}{2} h and k lines 
\citep{Au80,HS87,BT12,BT14} is confirmed by these new calculations in the 
presence of a magnetic field.

Secondly, M-O effects are found to be responsible for the appearance 
of important levels of \emph{broadband} Stokes-$U$ polarization. This 
result contrasts the common belief that magnetic fields produce 
significant polarization only in the line core, and in particular that 
the manifestation of M-O effects requires the presence of strong 
magnetic fields \citep[e.g.,][]{SL87}.
The fact that M-O effects are so outstanding in the polarization
profiles of the \ion{Mg}{2} h--k lines is due to the peculiar
combination of a strong opacity in the far wings and a significant 
level of scattering polarization (${\sim}2$\%; see Figure~\ref{fig:both}) 
induced by radiation anisotropy, which are produced in the adopted 
atmospheric model.

In our two-term model atom, these M-O effects induce the appearance of 
a polarization signal that encompasses the entire spectral range of 
the h--k doublet (spanning several nm; see Figures~\ref{fig:both} 
and \ref{fig:both.100}), 
and which is dominated by the signature of quantum interferences 
in the upper term of the \ion{Mg}{2} atomic model. 

A very remarkable result is the manifestation of these polarization
transfer effects already for relatively weak fields---in the modeled 
case of the \ion{Mg}{2} h and k lines, for field strengths of only a 
few gauss. This opens a completely new diagnostic window for the 
magnetism of the quiet-Sun upper atmosphere, since these effects 
should be detectable also in other notable chromospheric lines, such 
as the \ion{Ca}{2} H--K doublet around 395\,nm, the \ion{Na}{1} D-doublet 
around 590\,nm, and the \ion{H}{1} H$\alpha$ line at 653\,nm.

In the context of this work, the presence of broadband signals in the 
Stokes $Q$ and $U$ polarizations of the \ion{Mg}{2} h--k doublet make 
this line set a very attractive and potentially powerful diagnostic 
for synoptic magnetic studies of the solar chromosphere and upper 
photosphere. The relatively large amplitude of the signals over a 
spectral range spanning many Doppler widths should facilitate the design 
of high-throughput and fast cadence imaging polarimeters, which could 
rely on relatively low polarimetric sensitivity and spectral resolution,
at least for the diagnosis of the upper photosphere and lower
chromosphere, where these broadband signatures are produced.
The inner cores of the k and h lines, where the signatures of the 
Hanle and Zeeman effects dominate the polarization signal, probe
instead the low transition region (above ${\sim}2000$\,km in the 
FAL-C model).
Hence, \emph{the broadband Stokes profiles of the \ion{Mg}{2} h and k lines
offer an opportunity to study simultaneously the magnetic structure at 
the base and the top of the chromosphere.}

Our simulations show that the amplitude of the Hanle-effect polarization 
in the core of the \ion{Mg}{2} k-line can be as large as ${\sim}1$\%, 
and therefore relatively easy to detect using narrowband (${\sim}0.25$\AA) 
filter polarimeters.
In fact, a systematic study of the variation of narrowband-integrated 
Stokes $Q$ and $U$ polarizations as a function of the vector magnetic 
field should be conducted in order to assess the feasibility of filter-based, 
full-disk polarimeters for the \ion{Mg}{2} h--k doublet.

Finally, \emph{the magnetograph formula applied to the Stokes $V$ 
profiles of the \ion{Mg}{2} h--k doublet retains its diagnostic value 
as a proxy of the magnetism of the low transition region,} 
although our modeling also shows that its applicability breaks down
when PRD effects in the line formation region become important.
This conclusion reinforces the relevance of these lines 
for the diagnosis of chromospheric magnetic fields, and in particular
it provides a direct and inexpensive tool for the quick-look inversion of
large spectro-polarimetric datasets of chromospheric lines, providing
additional evidence to the importance and feasibility of full-disk 
observations of the solar chromosphere at these wavelengths.

\begin{acknowledgments}
The authors are grateful to J.~O.\ Stenflo (ETH Z\"urich) for 
internally reviewing the manuscript during a scientific visit to HAO,
and for helpful comments. The authors also thank E.\ Alsina Ballester, 
J.\ Trujillo Bueno (both IAC), and L.\ Belluzzi (IRSOL), who have been working
on a similar problem within the two-level atom formalism, for helpful 
discussions. We are deeply indebted to the referee for raising pointed
questions that have helped us improving the presentation of this letter.
\end{acknowledgments}

\end{document}